\documentclass[iop]{emulateapj}  

\usepackage{ulem}

\newcommand\aov{\ifmmode{\alpha_{\rm ov}}\else $\alpha_{\rm ov}$\fi}
\newcommand\fov{\ifmmode{f_{\rm ov}}\else $f_{\rm ov}$\fi}
\newcommand\amix{\ifmmode{\alpha_{\rm MLT}}\else $\alpha_{\rm MLT}$\fi}

\begin{document} 

\submitted{}

\title{Correction of the orbital mass of double galaxies estimation}

\author{
L. Belyaeva\altaffilmark{1} and
S. Parnovsky\altaffilmark{2}
}

\altaffiltext{1}{National Taras Shevchenko University of Kyiv, Kyiv, Ukraine}

\altaffiltext{2}{Astronomical observatory of the Taras Shevchenko National University of Kyiv, Kyiv, Ukraine}

\begin{abstract}
We obtain a more accurate statistical estimation of the mass of double galaxies moving in circular orbits, including confidence intervals for different confidence levels.
\end{abstract}


\section{Introduction}
\label{sec:introduction}

Determination of the mass of galaxies is one of the most difficult problems in extragalactic astronomy. One of the methods of estimating the mass of double galaxies is associated with the assumption of the motion of galaxies in a closed Keplerian orbit. The method of determining the mass of double galaxies was developed by Page \citep{Page:3, Page:1, Page:4, Page:2}. Later this approach was improved in the works of Holmberg \citep{Holmberg:1}, Karachentsev and Shcherbanovsky \citep{Karach:4}, Noreldlinger \citep{Noerdlinger:1}, Karachentsev \citep{Karach:3, Karach:2, Karach:1}.

\section{Determination of the orbital mass}

Karachentsev I.D. \citep{Karach:1} considers the physical pair of galaxies that carry orbital motion around a common center of mass. In the simplest case, we are dealing with a circular orbit for which according to Kepler's third law the total mass of galaxies is determined by the formula:

\begin{equation}\label{eq:eq1}
M = \frac{R_p(\Delta V_r)^2}{G \eta}
\end{equation}

\begin{equation}\label{eq:eq2}
M = K \frac{R_p(\Delta V_r)^2}{G} \; \; \; \; \; K = \eta^{-1}
\end{equation}
where $R_p$ -- projection of the distance between galaxies on the picture plane, $\Delta V_r$ -- relative radial velocity, $\eta$ -- a geometrical projection factor that has the form:

\begin{equation}\label{eq:eq3}
\eta = \sin^2 i \cos^2 \Omega (1 - \sin^2 i \sin^2 \Omega)^{\frac{1}{2}}
\end{equation}

$R_p$ are $\Delta V_r$ determined from observation but for an individual galaxy the geometric factor $\eta$ cannot be determined, therefore, statistical method of evaluation is used. An assumption is made about the random position of galaxies in relation to the line of sight. Then, the simultaneous distribution of the random quantities $i$ and $\Omega$ in this case has the form:

\begin{equation}\label{eq:eq4}
p_k (i,\Omega) = \frac{2}{\pi} \sin i \; \; \; \; \; [0 < i <\frac{\pi}{2}; 0 < \Omega < \frac{\pi}{2}]
\end{equation}

Further in the work \citep{Karach:1} Karachentsev proposed to use the expected value of a geometrical projection factor $<\eta> = \frac{3\pi}{32}$. Therefore, we get an estimate of the coefficient that is being used at the moment:

\begin{equation}\label{eq:eq5}
K_0 = 	\frac{3\pi}{32}
\end{equation}

\begin{equation}\label{eq:eq6}
M = 	\frac{32}{3\pi G}R_p (\Delta V_r)^2
\end{equation}

\section{Changing the existing estimation of the orbital mass}

Generally speaking, $<\frac{1}{\eta}> \neq \frac{1}{<\eta>}$, so it is interesting to investigate the distribution of $K$. If we try to calculate the expected value of $K$, we can see that the integral diverges and therefore no expected value exists. In such cases, the median is used as an estimate of the central distribution tendency \citep{Demidenko:1}. The median is considered a robust estimate \citep{Demidenko:1} and can be quantified numerically.

Using computer simulation, a median of the distribution was calculated, which is proposed to be used to estimate the total mass of galaxies. Then the new estimation is $1.54$ times more than (\ref{eq:eq4}) and looks like (\ref{eq:eq2}) with

\begin{equation}\label{eq:eq7}
K = 1.54 K_0
\end{equation}

Of course, estimation is still quite rough. For some orbits, we can get a mass much less than the real one. In view of this, other quantiles of distribution were also calculated (results are shown in Table \ref{tab:quantiles}). Table \ref{tab:quantiles} contains confidence intervals for different confidence probabilities and clearly illustrates in what limits the mass of double galaxies can vary.

\begin{table}[h]
\caption{Values of quantiles of the ratio $\frac{K}{K_0}$ distribution.\label{tab:quantiles}}
\centering
\begin{tabular}{lcc}
\hline
Probability $q$,\% & Quantile $\alpha_q$ \\
\hline
$50$       &   1.54    \\ 
$84.13$    &   20.26   \\ 
$15.87$    &   0.45    \\ 
$97.72$    &   1227    \\ 
$2.28$     &   0.31    \\ 
$95$       &   238     \\
$5$        &   0.33    \\ 
$97.5$     &   1015    \\ 
$2.5$      &   0.31   \\
\hline
\end{tabular}
\end{table} 

So, the lower and upper limits of the $1 \sigma$ confidence interval are $0.45$ and $20.26$ respectively and we propose to use the factor $K = 1.5^{+18.7}_{-1.1}$ in the equation (\ref{eq:eq2}). This confidence interval is very asymmetrical, so an estimation of its boundary based on the statistical distribution of $K$ is very useful. Estimation of the confidence intervals limits for some popular confidence levels one can find in Tab. 1.

\section{Conclusion}
The method of measuring the mass of double galaxies was considered. The use of the mass distribution median is proposed instead of the inversed expected value of a geometrical projection factor. As a result, we propose some corrections to the formula that was used for years. In addition, the confidence intervals for different confidence probabilities were calculated to estimate its accuracy.\\

\end{document}